\documentclass[9pt]{IEEEtran}
\IEEEoverridecommandlockouts
\usepackage{soul}
\usepackage{cite}
\usepackage{amsmath,amssymb,amsfonts}
\usepackage{algorithmic}
\usepackage{graphicx}
\usepackage{tabularx, booktabs}
\usepackage{microtype}
\usepackage{textcomp}
\usepackage{url}
\usepackage{xcolor}
\def\BibTeX{{\rm B\kern-.05em{\sc i\kern-.025em b}\kern-.08em
    T\kern-.1667em\lower.7ex\hbox{E}\kern-.125emX}}
\begin{document}

% \title{Conference Paper Title*\\
% {\footnotesize \textsuperscript{*}Note: Sub-titles are not captured for https://ieeexplore.ieee.org  and

% }
\title{LLaQo: Towards a Query-Based Coach in \\ Expressive Music Performance Assessment
}

\author{
\IEEEauthorblockN{Huan Zhang
\thanks{Work done during internship at Sony CSL.}
$^{\star}$ \qquad Vincent K.M. Cheung$^{\dagger}$ \qquad  Hayato Nishioka$^{\dagger}$ \qquad  Simon Dixon$^{\star}$ \qquad  Shinichi Furuya$^{\dagger}$}\\
\IEEEauthorblockA{$^{\star}$ Centre for Digital Music, Queen Mary University of London, UK \\
$^{\dagger}$}Sony Computer Science Laboratories, Tokyo, Japan \\
\texttt{huan.zhang@qmul.ac.uk}
}

% \author{

% \IEEEauthorblockN{1\textsuperscript{st} Given Name Surname}
% \and
% \and
% \and
% \and
% \and

\maketitle

\begin{abstract}
Research in music understanding has extensively explored composition-level attributes such as key, genre, and instrumentation through advanced representations, leading to cross-modal applications using large language models. However, aspects of musical performance such as stylistic expression and technique remain underexplored, along with the potential of using large language models to enhance educational outcomes with customized feedback. To bridge this gap, we introduce LLaQo, a \textbf{L}arge \textbf{La}nguage \textbf{Q}uery-based music c\textbf{o}ach that leverages audio language modeling to provide detailed and formative assessments of music performances. We also introduce instruction-tuned query-response datasets that cover a variety of performance dimensions from pitch accuracy to articulation, as well as contextual performance understanding (such as difficulty and performance techniques). Utilizing AudioMAE encoder and Vicuna-7b LLM backend, our model achieved state-of-the-art (SOTA) results in predicting teachers' performance ratings, as well as in identifying piece difficulty and playing techniques. Textual responses from LLaQo was moreover rated significantly higher compared to other baseline models in a user study using audio-text matching. Our proposed model can thus provide informative answers to open-ended questions related to musical performance from audio data.

\end{abstract}


\section{Introduction}\label{sec:introduction}

In the domain of Music Information Retrieval, the majority of music understanding research has focused on composition-level attributes such as key, tempo, genre, and instrumentation. 
\cite{Korzeniowski2017EndNetwork, Pons2018End-to-endScale, FloresGarcia2021LeveragingRecognition}
These attributes are not only tagged individually via end-to-end approaches but have also been the focus of foundation models and unified music representations. Recent advances further extended to cross-modal understanding of music, which enabled tasks such as  music captioning \cite{Liu2024MusicCaptioning, Doh2023LP-MusicCapsCaptioning} and even music reasoning \cite{Gardner2024LLarkMusic} with the help of large language models (LLMs). Despite these technological advances, the nuanced aspects of music performance — such as interpreting complex techniques, recognizing stylistic differences, and evaluating performance quality — remain underexplored \cite{Zhang2024FromPiano}. 

On the other hand, traditional systems for music performance assessment \cite{Lerch2019MusicSurvey} have primarily relied on quantitative metrics that provide summative feedback, often in the form of scores or binary judgments \cite{Zhang2021LearnAssessment, Huang2020Score-informedAssessment, Seshadri2021ImprovingLearning}. Such approaches, while beneficial for certain applications, fall short in educational contexts where formative, detailed and instructional feedback is essential for student development \cite{Eremenko2020PerformanceLearning, Morsi2024SimulatingLearning}. Meanwhile, performance assessment encompasses multiple dimensions including rhythm accuracy, tone production, and expressive dynamics, which makes semantic descriptors particularly appropriate. Thus, we introduce \textbf{LLaQo} (Large Language Query-based Coach), a pioneering approach in the domain of expressive performance assessment that leverages the capabilities of LLMs to analyze and assess music performances at a granular level, providing interpretative guidance for MIR-supported music pedagogy. Constructed with AudioMAE for audio embeddings, Vicuna-7b language backend along with a Q-former aligner, our model is trained on instruction-tuned data specific to performance assessment. Our contribution\footnote{Code will be made available at: \url{https://github.com/anusfoil/LLaQo}, Data is demonstrated in project page: \url{https://bit.ly/3XjUOuX}} can be summarized as follows: 
\begin{enumerate}
    \item We present the first LLM-supported music coach, addressing performance understanding tasks including difficulty analysis, technique analysis, composer recognition, and most importantly, performance assessment. 
    \item We compiled and instruction-tuned a diverse collection of publicly available performance-understanding datasets, and standardized them into query-response pairs. Additionally, we contribute a newly recorded dataset, \textit{NeuroPiano}, with detailed annotations on 12 performance dimensions.
    \item Our model demonstrated superior performance in objective and subjective evaluation tasks over existing audio-language model baselines in performance assessment, as well as difficulty and technique prediction.
\end{enumerate}

% In the next section we review related works; In section 3, we formulate the task and discuss our data compilation. In section 4, we talk about methodology and model architectures. In section 5, we show how is the model being evaluated. 

\begin{figure*}
    \centering
    \includegraphics[width=\linewidth, trim={0.2cm 2cm 0.2cm 3.5cm}]{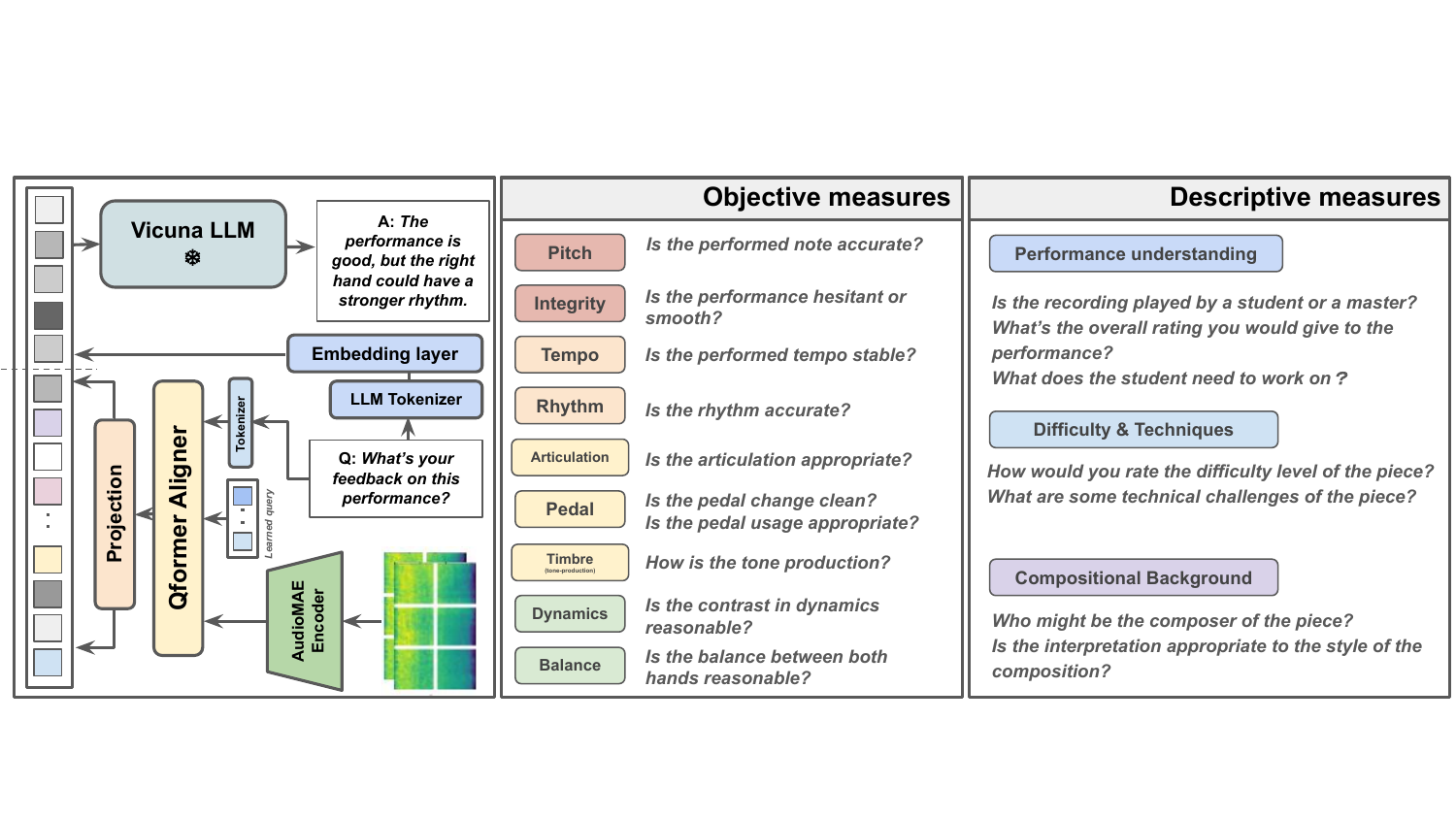}
    \caption{Overall framework of our model, as well as the piano assessment rubric used to construct query-answer pairs for instruction-tuning in each dataset. 
    % \textcolor{red}{Objective/Descirptive measureS -- add 's'; Is the contrast in dynamics reasonable?, Is the balance between both hands reasonable? What are some technical challenges of the piece?}
    }
    \label{fig:framework_and_instruction}
\end{figure*}

\section{Related Work}

The integration of large language models with audio signal processing has been overviewed in Latif et al. \cite{Latif2023SparksOutlook} and Wu et al. \cite{Wu2024TowardsOverview}. Pioneering audio language models (ALM) such as
% AudioPALM \cite{Rubenstein2023AudioPaLMListen}, 
Pengi \cite{Deshmukh2023PengiTasks}, LTU \cite{Gong2023JointUnderstanding} and SALMONN \cite{Tang2024SALMONNModels} bridge audio and speech perception with natural language reasoning and specializes in tasks from speech recognition to audio reasoning. Such ALMs are usually composed of a frozen audio encoder (CLAP\cite{Elizalde2023CLAPSupervision}, Whisper\cite{Radford2023RobustSupervision}) and a frozen causal language model (LLaMA\cite{Touvron2023LLaMAModels}, GPT2) as backend, and a trained connection or mapping module that project the input audio embeddings into the text token space. Given an  input audio-text pair, the models can generate free-form text to address open-ended tasks and close-ended tasks. In the specific branch of music language understanding, however, research has mainly focused on music context question answering \cite{Gardner2024LLarkMusic} and captioning \cite{Liu2024MusicCaptioning}, where models are limited to answering questions on compositional aspects such as genre, mood or instrumentation. On the other hand, pure LLMs showed poor musical understanding and are only restricted to answering questions related to the contextual background behind a musical piece \cite{Li2024MusicModels, Ma2024FoundationSurvey}. Models with adequate musical performance understanding are therefore still lacking \cite{Zhang2024FromPiano}.

Another challenge in the field of music understanding relates to the availability and quality of data. Efforts have been made to generate natural language descriptions from large-scale tag datasets through language models, as seen in music captioning initiatives \cite{Doh2023LP-MusicCapsCaptioning}. Additionally, instruction-tuning has been applied to unify disparate music data annotations for fine-tuning LLMs \cite{Gardner2024LLarkMusic}. In the realm of music education, however, acquiring precise descriptors for performance nuances proves difficult \cite{Morsi2024SimulatingLearning}, whether through automatic means or manual annotation. This task is further complicated by the inherent perceptural subjectivity, which can lead to bias and inconsistency in human feedback \cite{Jiang2023ExpertFeedback, Lerch2019MusicSurvey}. Despite these challenges, the field of automatic music assessment has seen considerable benefits from leveraging formative text feedback \cite{Eremenko2020PerformanceLearning}, which provides a richer alternative to traditional methods that rely on simplistic summative scoring systems \cite{Zhang2021LearnAssessment, Huang2020Score-informedAssessment, Seshadri2021ImprovingLearning}.

\section{Instruction tuning dataset on \\ music education}

In Table~\ref{tab:datasets_overview}, we describe the  instruction-tuned expressive performance datasets used to train our framework. Several types of data were utilized:  

The first consisted of teachers' free-response feedback \footnote{The CROCUS dataset is translated from Japanese to English using DeepL.} \cite{Jiang2023ExpertFeedback, Matsubara2021CROCUSCritiques} that provides a holistic comment on a performance, covering various performance dimensions or practice suggestions. This data was transformed into query-response pairs using \texttt{GPT4} according to the instruction of an assessment rubric in Figure~\ref{fig:framework_and_instruction}. For each piece of feedback (around 30 to 50 words), we generated three to five QA pairs. 

The second comprised data with performance-related annotations \cite{Sarasua2019DatasetGestures, Li2023SiameseAssessment}, such as a specific articulation or tempo. In this case, queries were formed based on their performance attributes, with examples such as \textit{Is the playing using legato or staccato articulation?}  

The third consisted of single-rating datasets \cite{Parmar2021PianoAssessment, Wang2021Audio-basedMechanism} that come with a summative rating. The query was designed to ask directly about the rating, such as \textit{How would you rate the overall performance, in a scale of 10?} 

The last data type comprised performance understanding data \cite{Zhang2024FromPiano, Ramoneda2023CombiningClassification} that do not address the quality of a performance, but on contextual information that is ideally obtained by a teacher or student, such as \textit{What's the most difficult technique in this passage?} or \textit{What is the composer and stylistic period of this piece?} 
In total, we aggregated around 130 hours of recording and 34k QA pairs for training.

For this project, we also recorded a student pianist dataset with detailed teacher annotations (the \textit{NeuroPiano} dataset). It consists of short technique snippets including scales, stepwise chords, and third dyads performed by students\footnote{From the NeuroPiano Music academy: \url{www.neuropiano.org}}. Based on the rubric in Figure~\ref{fig:framework_and_instruction}, we formulated 12 questions, ranging from tone production to hand balance, in which teachers provided textual feedback and a rating score on a scale from 1 to 6. Note that the original annotations are in Japanese, in which we used DeepL to translate to English, and performed a round of manual correction. The dataset is further spitted by half for training, with the remaining half used for objective and subjective evaluation described in Section~\ref{sec:eval}.

\begin{table*}
    \small
    \centering
    \begin{tabularx}{\linewidth}{p{3cm} p{1.2cm} p{1.2cm} p{2cm} p{3.5cm} X}
        \toprule
        \textit{Dataset}  & \textit{Recordings} & \textit{QA pairs} & \textit{Level} & \textit{Feedback or Annotation} & \textit{Repertoire} \\ 
        \midrule
        Expert-Novice (\cite{Jiang2023ExpertFeedback}) & 83 & 2.4k & beginner & Full verbal feedback and ratings & Pop song arrangements \\ [0.5ex]
        % \addlinespace 
        CROCUS-piano (\cite{Matsubara2021CROCUSCritiques}) & 22 & 616 & intermediate, advanced &  Verbal feedback  & Western classical concert repertoire \\ [0.5ex]
        % \addlinespace 
        \textit{Con Espressione} (\cite{Cancino-Chacon2020OnGame})  & 50 & 4.3k & master, MIDI  & Description of expressive character  & Western classical concert repertoire  \\[0.5ex]
        % \addlinespace 
        Expressive Musical Gestures (\cite{Sarasua2019DatasetGestures})  & 106 & 845  & intermediate & Labels of performance instruction (e.g. slow, staccato) & `Träumerei' by R. Schumann  \\[1ex]
        % \addlinespace 
        Burgmuller (\cite{Morsi2023SoundsPerformances})  & 25 & 114 & intermediate & Annotation of error regions & Burgmuller Etudes \\[0.5ex]
        % \addlinespace 
        Music Shape Dataset (\cite{Li2023SiameseAssessment})  & 2.3K & 11k  & intermediate &  Annotations of dynamics and articulation changes & Schmitt exercise \\[0.5ex]
        % \addlinespace 
        PISA (\cite{Parmar2021PianoAssessment}) & 61 & 256  & beginner-intermediate & Single rating score  & Folk songs arrangements, easy classical pieces \\[0.5ex]
        % \addlinespace 
        YCU-PPE (\cite{Wang2021Audio-basedMechanism}) & 2.6K & 7.7k  & beginner & Rating score from 3 judges  & Chinese folk song arrangements \\[0.5ex]
        \addlinespace 
        \hline
        \addlinespace         
        PLD-expertise (\cite{Zhang2024FromPiano, Zhang2022ATEPPPerformance}) & 1.4k & 2k & beginner and advanced & Performing skill level based on the uploader's profile & Western classical concert repertoire  \\[0.5ex]
        % \addlinespace        
        PLD-techniques (\cite{Zhang2024FromPiano}) & 223 & 223 & mixed & Piano Technique labeling (e.g. arpeggio, octave) & Music segments of common piano techniques such as scales, ornaments, repeated notes, etc. \\[0.5ex]
        % \addlinespace        
        CIPI (\cite{Ramoneda2023CombiningClassification}) & 736 & 2.2k  & intermediate to advanced & Henle Difficulty labeling on a scale of 9 & Western classical concert repertoire \\
        % \addlinespace        
        \textbf{NeuroPiano} (ours) & 104 & 3.3k  & advanced & text response on 12 specific aspects including articulation and tempo, as well as rating score on a scale of 6 & Short technique snippets including scales, arpeggios, dyads, blocked chords, octaves \\
        \bottomrule    
    \end{tabularx}
    \vspace{0.1cm}
    \caption{Public datasets related to performance assessment and performance understanding, instruction-tuned for training our model.}
    \label{tab:datasets_overview}
\end{table*}

\section{Model architecture and training}

The overall structure of LLaQo is based on the APT-LLM \cite{Liang2023AcousticCapabilities} framework, which originates from BLIP-2 framework, as shown in Fig~\ref{fig:framework_and_instruction}. It is comprised of an audio encoder, an audio-language aligner, and a large language model.  We use Audio-MAE\cite{Hu2022MaskedListen}, a 12-layer transformer encoder that learns to reconstruct randomly-masked spectrogram patches during training, as the audio encoder. The output feature map from the penultimate block of Audio-MAE encodes fine-grained patterns that's essential for performance understanding. 

The audio-language aligner connects the audio encoder to the frozen language model. It takes in a text prompt together with audio feature maps extracted by the audio encoder, and produces a fixed number of acoustic embeddings. Following a Query Transformer (Q-former) architecture, four transformer blocks constitute our audio aligner where 32 trainable query embeddings attend to the input text tokens and extract relevant information from the audio feature maps. 

The language model predicts the output text by considering the previous generated texts and the input audio-text tokens. When interleaving audio and text tokens, each audio clip is appended with a learnable token $\langle$AUDIO$\rangle$ to indicate the beginning of audio tokens. As shown in Figure~\ref{fig:framework_and_instruction}, the query text pass through both the Q-former tokenizer and LLM's tokenizer and its word embedding layer. We parameterize the language model by Vicuna-7b\cite{Zheng2023JudgingArena}. 

\subsection{Learning Objective}
Consider an audio-text pair $(a, t)$ and their response $g$. The audio input $a$ is transformed into a sequence of embeddings $\mathbf{X}_{\text{audio}}$ using the encoder $\mathcal{A}_{\phi}$ and aligner $\mathcal{M}_{\theta}$. These embeddings are then concatenated with text embeddings $\mathbf{X}_{\text{text}}$, derived from input text $t$ using the embedding layer $\mathbf{W}_{\psi}$ to form the combined sequence $\mathbf{X}_{\text{audio;text}}$. More precisely, the embeddings are concatenated as follows:
\begin{equation}
\mathbf{X}_{\text{audio;text}} = C(\mathcal{M}_{\theta}(\mathcal{A}_{\phi}(a)), \mathbf{W}_{\psi}(t)),
\end{equation}
where $C$ denotes the concatenation function.

The learning objective $\mathcal{L}$ of our model is to optimize the parameters $\phi$, $\theta$, and $\psi$ to maximize its ability to predict the next token based on the concatenated embeddings by minimizing cross-entropy loss:
\begin{equation}
\mathcal{L} = 
\sum_{i=L+1}^{L+|g|} \text{log} p_{\phi, \theta, \psi}(\mathbf{X}_i|\mathbf{X}_{\text{audio;text}}, \mathbf{X}_{\text{pred},<i}),
\label{eq:prediction}
\end{equation}
where $L$ is the length of $\mathbf{X}_{\text{audio;text}}$. This formulation allows us to directly leverage the audio-text context to predict subsequent tokens.

% \vspace{-10pt}

\subsection{Experiment setup}

\noindent \textbf{Pretraining}: We first pretrain the audio aligner module to bridge the audio modality and the text modality. The Qformer audio aligner is trained with audio-text pairs from AudioSet and WavCaps, with the triplet objective of audio-text matching, audio-grounded text generation, and audio-text contrastive as described in \cite{Liang2023AcousticCapabilities}. This step is performed on four NVIDIA A100 (40G) and trained for 120 hours. 

\noindent \textbf{General audio multi-task finetuning}: The first-stage finetuning trains the whole pipeline as described in the previous section. It is performed on general audio tasks such as \textit{audio tagging, audio captioning, audio question answering,} facilitated by datasets including AudioSet, WavCaps, and Clotho. Furthermore, it learns from multiple audio clips through few-shot audio classification and natural language audio reasoning, where it predicts sound event labels by juxtaposing more than one audio clips with input text. This step is performed on four NVIDIA A100 (40G) and trained for 120 hours. 

\noindent \textbf{LLaQo data finetuning}: The final round of finetuning is trained with intruction-tuned datasets in Table~\ref{tab:datasets_overview}, with $(a, t, g)$ triplets as described in Equation~\ref{eq:prediction}. This last step is performed on one NVIDIA H100 (80G) GPU and trained for 120 hours. 

Adam optimiser was used for all model training described above. We applied linear warmup strategy in the first 2K steps and used a cosine annealing learning rate of $5\times10^5$. For the AudioMAE input, we resampled all data into 32kHz and computed Kaldi filterbank with 128 mels and a frameshift of 10$ms$. Due to resource limitation, our input is limited to a filterbank of length 4096 (40 seconds).

\begin{table*}[ht]
    \centering
    \normalsize
    \begin{tabular}{lcccccccccc}
        \toprule
        & \multicolumn{6}{c}{Performance assessment} & \multicolumn{2}{c}{Difficulty} & \multicolumn{1}{c}{Technique} \\
        \cmidrule(lr){2-7} \cmidrule(lr){8-9} \cmidrule(lr){10-10}
        Model & PFR$\uparrow$ & MAE$\downarrow$ & SS$\uparrow$ & B-U$\uparrow$ & SPICE$\uparrow$ & BERT-S$\uparrow$  & $Acc_0\uparrow$ & $Acc_1\uparrow$ & $Acc_0\uparrow$ \\
        \midrule
        LLaQo & \textbf{0.99$\pm$0.02} & \textbf{0.97$\pm$0.10} & \textbf{0.52$\pm$0.08} & \textbf{0.09$\pm$0.02} & \textbf{0.12$\pm$0.04} & \textbf{0.88$\pm$0.07} & \textbf{0.21$\pm$0.10} & \textbf{0.55$\pm$0.15} & \textbf{0.37$\pm$0.12} \\
        LTU & 0.98$\pm$0.01 & 1.39$\pm$0.08 & 0.32$\pm$0.05 & 0.06$\pm$0.03 & 0.07$\pm$0.10 & 0.87$\pm$0.12 & 0.16$\pm$0.09 & 0.34$\pm$0.13 & 0.16$\pm$0.07 \\
        MU-LLaMa & 0.66$\pm$0.07 & 1.81$\pm$0.15 & 0.40$\pm$0.10 & 0.06$\pm$0.05 & 0.12$\pm$0.06 & 0.88$\pm$0.09 & 0.11$\pm$0.12 & 0.26$\pm$0.18 & 0.20$\pm$0.10 \\
        \bottomrule
    \end{tabular}
    \label{tab:obj_eval}
    \vspace{0.1cm}
    \caption{Comparison of models on the metrics of the evaluated content. Numbers indicate mean and standard deviation. $\uparrow$ indicates larger value for better performance, and vice versa. }
\end{table*}

\section{Evaluation and discussions}
\label{sec:eval}

\subsection{Compared baselines}

For all tasks, we compared our model against two open-source audio-language models capable of answering audio QA tasks, namely \noindent \textbf{Listen, Think and Understand (LTU-AS)} \cite{Gong2023JointUnderstanding}, which is an improvement of the LTU \cite{Gong2023ListenUnderstand} that incorporates Whisper\cite{Radford2023RobustSupervision} and low-rank adaptors to the language model, as well as \textbf{Music Understanding LLaMa (MuLLaMA)} \cite{Liu2024MusicCaptioning}, which is trained on a generated MusicQA dataset and utilizes MERT encoder for audio representation to answer music-related questions and generate captions for tracks.

\subsection{Objective evaluation}
Given an audio input, we first evaluated the performance of each model in terms of its ability to answer open and closed questions regarding \textit{Performance assessment}, \textit{Difficulty}, and \textit{Technique}. We performed three iterations of objective evaluation to ensure that the reported results were stable.

To evaluate Performance assessment, each model was required to provide open responses to 12 questions related to the quality of a performance (see Figure~\ref{fig:framework_and_instruction}), as well as an integer score from 1 to 6 (e.g., with the prompt \textit{`How would you rate the consistency of the tempo on a scale of 1 to 6?'}) on the evaluation set of the NeuroPiano dataset. Model performance on open questions was assessed in terms of semantic validity and given in terms of common NLP metrics, namely \textit{BLEU} (B-U) (an average of BLEU1, BLEU2, BLEU3 and BLEU4), \textit{SPICE} \cite{Anderson2016SPICEevaluation} and \textit{BERT-Score} (BERT-S). We also measured \textit{sentiment similarity} (SS), which is quantified by the cosine similarity of a 5-class sentiment vector between teachers' feedback and model output as given by a sentiment analysis model\footnote{\url{https://huggingface.co/nlptown/bert-base-multilingual-uncased-sentiment}}. Performance on closed questions was quantified using \textit{mean absolute error} (MAE), which measures the deviation from the predicted rating with the teacher rating, and \textit{prompt following rate} (PFR), which defines the ratio in which the model actually produced a numerical rating. Given that baseline models LTU and Mu-LLaMa are not specifically trained for numerical predictions, we allowed a maximum of 5 attempts for them to generate a numerical rating.   

To evaluate Difficulty and Technique, we followed the same procedure as in \cite{Zhang2024FromPiano}. For Difficulty, each model was required to produce an integer score from 1 to 9 given the prompt \textit{`How would you rate the difficulty of the piece on a scale of 9?'} on the CIPI dataset \cite{Ramoneda2023CombiningClassification}. Model performance was quantified in terms of proportion of exact matches ($Acc_0$), and proportion of matches that deviate by at most 1 ($Acc_1$). For Technique, each model was required to select 1 of 7 playing techniques in response to the prompt \textit{`What's the most salient technique used in this recording? Choose from trills, octaves...'} on the PLD-techniques dataset \cite{Zhang2024FromPiano}, and evaluated in terms of $Acc_0$. 

The results, presented in Table~\ref{tab:obj_eval}, reveal that our proposed LLaQo model achieved the best performance amongst all models with regards to all objective metrics. Numerical predictions by LLaQo was within a MAE of less than 1 on a scale of 6, indicating a close alignment between its predictions and the teachers' ground truth. Also, while  LLaQo and LTU were able to consistently return numerical ratings, MU-LLaMa struggled, as shown by a low PFR score.

Generated repsonses by LLaQo also showed substantially higher sentiment similarity with the ground truth (0.52) compared to LTU and MU-LLaMA. However, the BLEU score was relatively low across all models compared to typical values of 0.2-0.3\cite{Liu2024MusicCaptioning} in music captioning tasks. This could be attributed to the expansive semantic space used by the teacher for feedback, where traditional NLP metrics might not have been suited for evaluating feedback responses in this context. Likewise, all models achieved similar BERT-S scores possible because all model responses remained on-topic. 

Lastly, while LLaQo surpassed the baseline models for difficulty and technique prediction by a wide margin, it is interesting to note that it did not outperform non-LLM based approaches on the same dataset (e.g. \cite{Ramoneda2023CombiningClassification} achieved $Acc_0$ = 0.39 in Difficulty, \cite{Zhang2024FromPiano} reported $Acc_0$ = 0.47 in Technique). 

\subsection{Subjective feedback evaluation}
We conducted a human-based evaluation using \textit{audio-text matching} \cite{Gardner2024LLarkMusic} to subjectively evaluate the quality of textual answers provided by each model. Given a query and audio, participants (20 music students and teachers) rated the response from each model on a 4-point scale in terms of their relevance to the question and alignment with the performance content. In particular, three categories of open-ended questions were matched:
\begin{enumerate}
    \item Feedback: \textit{What's your overall feedback on this performance?}, or \textit{How clean is the attack?}, or \textit{Is the dynamics change natural?} 
    \item Suggestion: \textit{What does the student need to work on? }
    \item Appreciation: \textit{How would you describe the emotional intent of the performance? }
\end{enumerate}
Participants rated 250 randomly sampled questions on Feedback and Suggestion from the \textit{NeuroPiano} dataset, as well as 250 questions on Feedback, Suggestion, and Appreciation from the Expert-Novice, CROCUS and YCU-PPE datasets. 

\begin{figure}
    \centering
    \includegraphics[width=\linewidth]{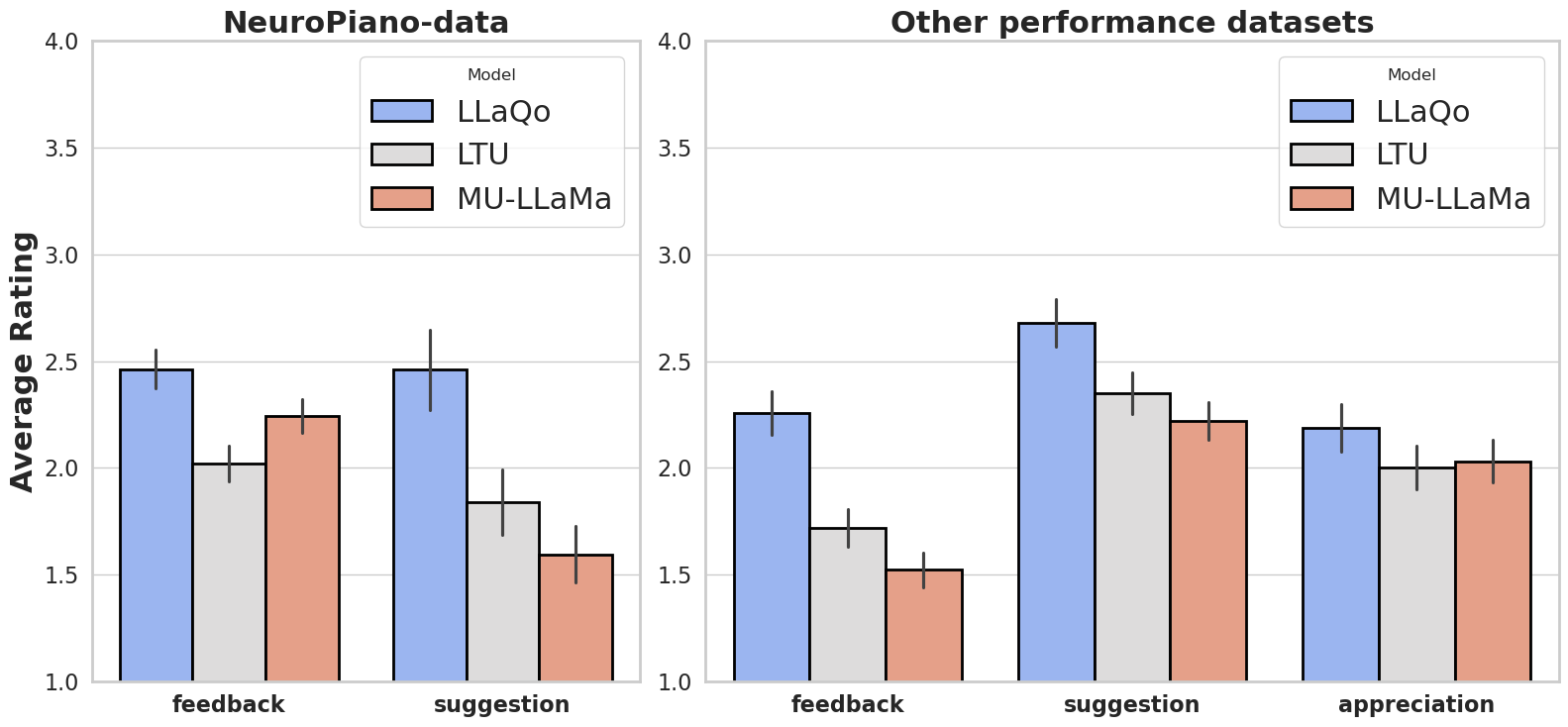}
    \caption{Audio-text matching subjective evaluation result of three models, with errors bars indicating standard error of the mean. 
    % \textcolor{red}{feedback suggestion appreciation order needs to be the same}
    }
    \label{fig:sbj_result}
\end{figure}

Results of the audio-text matching task are shown in Figure~\ref{fig:sbj_result}. In terms of Suggestions, participants rated our proposed LLaQo model significantly higher than LTU and MU-LLaMa in both \textit{NeuroPiano} and Other datasets (Mann-Whitney U-test, \textit{NeuroPiano}: $p(corrected) = 0.0274$ and $p = 0.0013$, respectively; Other: $p = 0.0382$ and $p = 0.0028$, respectively). Similarly, LLaQo was rated the highest for Feedback in both datasets, scoring significantly higher than LTU in both ($p = 7.2\times10^{-4}$ and $p = 1.9\times10{-4}$, respectively) and MU-LLaMa in Others ($p = 0.0028$). Lastly, participants gave highest ratings for LLaQo in questions related to Appreciation, though the differences were not statistically significant.

\section{Conclusion and Future Work}

This paper introduced the LLaQo model, designed for music-education related question answering, which is the first study that use Large Language Model to support music pedagogy. Our model offers performance feedback and suggestions, as well as information on performance techniques and piece difficulty, and has demonstrated superior performance in objective and subjective evaluation tasks over existing models such as LTU and MU-LLaMa. Additionally, we contributed a collection of instruction-tuned music question answering datasets, including the recorded and annotated in-house, and previously unreleased \textit{NeuroPiano} dataset.

Though our model is currently trained on piano sets, our framework can be easily extended towards other instruments, when large-scale assessment data become available. Future work should also consider timestamp positioning of feedback, as the ability to correctly locate an error, (e.g. the first measure, the last beat) is crucial for informative feedback. Advances in position-aware audio - or even musical score -representations may prove fruitful for this endeavor.

\section*{Acknowledgement}

This work is supported by the UKRI Centre for Doctoral Training in Artificial Intelligence and Music, funded by UK Research and Innovation [grant number EP/S022694/1], and Japan Science and Technology Agency CREST grant [number JPMJCR20D4]. The compute in this research utilised Queen Mary's Apocrita HPC facility, supported by QMUL Research-IT. \url{http://doi.org/10.5281/zenodo.438045}.

% \section*{References}
% \newpage
\bibliographystyle{IEEEtran}
\bibliography{ref}

\end{document}